\documentclass[fleqn,usenatbib]{mnras}

\usepackage{newtxtext,newtxmath}

\usepackage[T1]{fontenc}

\DeclareRobustCommand{\VAN}[3]{#2}
\let\VANthebibliography\thebibliography
\def\thebibliography{\DeclareRobustCommand{\VAN}[3]{##3}\VANthebibliography}

\usepackage{graphicx}	
\usepackage{amsmath}	

\title[The two-component magnetic field of M82]{Submillimetre observations of the two-component magnetic field in M82}

\author[K. Pattle et al.]{
Kate Pattle,$^{1}$\thanks{E-mail: katherine.pattle@nuigalway.ie} Walter Gear,$^{1}$, Matt Redman,$^{1}$, Matthew W. L. Smith,$^{2}$ and Jane Greaves$^{2}$ 
\\
$^{1}$Centre for Astronomy, Department of Physics, National University of Ireland Galway, University Road, Galway H91 TK33, Ireland\\
$^{2}$School of Physics and Astronomy, Cardiff University, The Parade, Cardiff CF24 3AA, United Kingdom 
}

\date{Accepted XXX. Received YYY; in original form ZZZ}

\pubyear{2021}

\begin{document}
\label{firstpage}
\pagerange{\pageref{firstpage}--\pageref{lastpage}}
\maketitle

\begin{abstract}
We observed the starburst galaxy M82 in 850$\mu$m polarised light with the POL-2 polarimeter on the James Clerk Maxwell Telescope (JCMT).
We interpret our observed polarisation geometry as tracing a two-component magnetic field: a poloidal component aligned with the galactic `superwind', extending to a height $\sim 350$\,pc above and below the central bar; and a spiral-arm-aligned, or possibly toroidal, component in the plane of the galaxy, which dominates the 850$\mu$m polarised light distribution at galactocentric radii $\gtrsim 2$\,kpc.
Comparison of our results with recent HAWC+ measurements of the field in the dust entrained by the M82 superwind suggests that the superwind breaks out from the central starburst at $\sim 350$\,pc above the plane of the galaxy.
\end{abstract}




\section{Introduction}

The magnetic fields observed in galaxies are thought to arise from the $\alpha\omega$-dynamo effect, in combination with a supernova-driven galactic wind \citep[e.g.][]{beck1996}.  This dynamo results in spiral galaxies typically showing a magnetic field running parallel to their spiral arms, likely wound up over several Gyr \citep[e.g.][]{jones2020}, while magnetic field transport in galactic winds and outflows is thought to be a key mechanism by which the intergalactic medium is magnetised \citep[e.g.][]{kronberg1999}.  The dynamo mechanism has been shown to persist through periods of starburst activity, and to be enhanced by tidal interactions \citep{moss2014}.

M82 is a nearby (3.5 Mpc; \citealt{jacobs2009}) edge-on starburst galaxy, interacting with its neighbour M81.  It has a bipolar `superwind' emanating from its central region \citep[e.g.][]{shopbell1998}.  M82 is classed as irregular, but nonetheless has a disc with a well-identified inclination angle of $76.9^{\circ}$ \citep{clark2018}, and a 1-kpc-long bar \citep{telesco1991} from which weak $m=1$ logarithmic spiral arms emanate \citep{mayya2005}.

The magnetic field of M82 is expected to consist of both a toroidal disc component and a poloidal wind component \citep[e.g.][]{jones2000}.  However, the brightness of the outflow-launching galactic nucleus, and the strength of the superwind, has made observing the planar component challenging \citep{jones2019}.

M82 was first observed in extinction polarisation by \citet{elvius1962}, and was further observed in optical and 1.65$\mu$m extinction polarisation by \citet{neininger1990} and \citet{jones2000} respectively.  These latter studies found a near-vertical field geometry in the galactic nucleus, inferred to trace the field in the superwind. 

M82 was observed at 850$\mu$m with SCUPOL on the James Clerk Maxwell Telescope (JCMT) by \citet{greaves2000}, who again found a near-vertical field geometry in the galactic nucleus, but saw a broadly elliptical magnetic field geometry in the outer galaxy, which they interpreted as a magnetic bubble driven by the superwind.  These data were reprocessed by \citet{matthews2009}, producing results consistent with those of \citet{greaves2000}, but in which the loop structure is less apparent.

M82 has been observed in radio polarisation (6 -- 22 cm) by \citet{adebahr2017}, who found a polarisation geometry consistent with a planar field in the inner part of the galaxy, which they interpret as tracing a field running along the galactic bar.

The magnetic field in M82 has recently been observed with HAWC+ at 53$\mu$m and 154$\mu$m \citep{jones2019}, finding that the near-vertical field geometry in the nucleus to extend into the galactic halo, inconsistent with the magnetic bubble model of \citet{greaves2000}.  The magnetic field lines seen by HAWC+ have been inferred to be open -- connecting the starburst core to the intergalactic medium, rather than creating a galactic fountain -- in a recent preprint \citep{lopezrodriguez2021}.

In this letter we present 850$\mu$m dust emission polarisation observations of M82 made using the POL-2 polarimeter mounted on the SCUBA-2 camera on the JCMT.
Section~\ref{sec:obs} describes the observations and the data reduction process.  Section~\ref{sec:results} describes our results.  Section~\ref{sec:discussion} compares our observations to previous measurements, and interprets our results.  Our conclusions are presented in Section~\ref{sec:summary}. 

\section{Observations}
\label{sec:obs}

\begin{figure*}
    \centering
    \includegraphics[width=\textwidth]{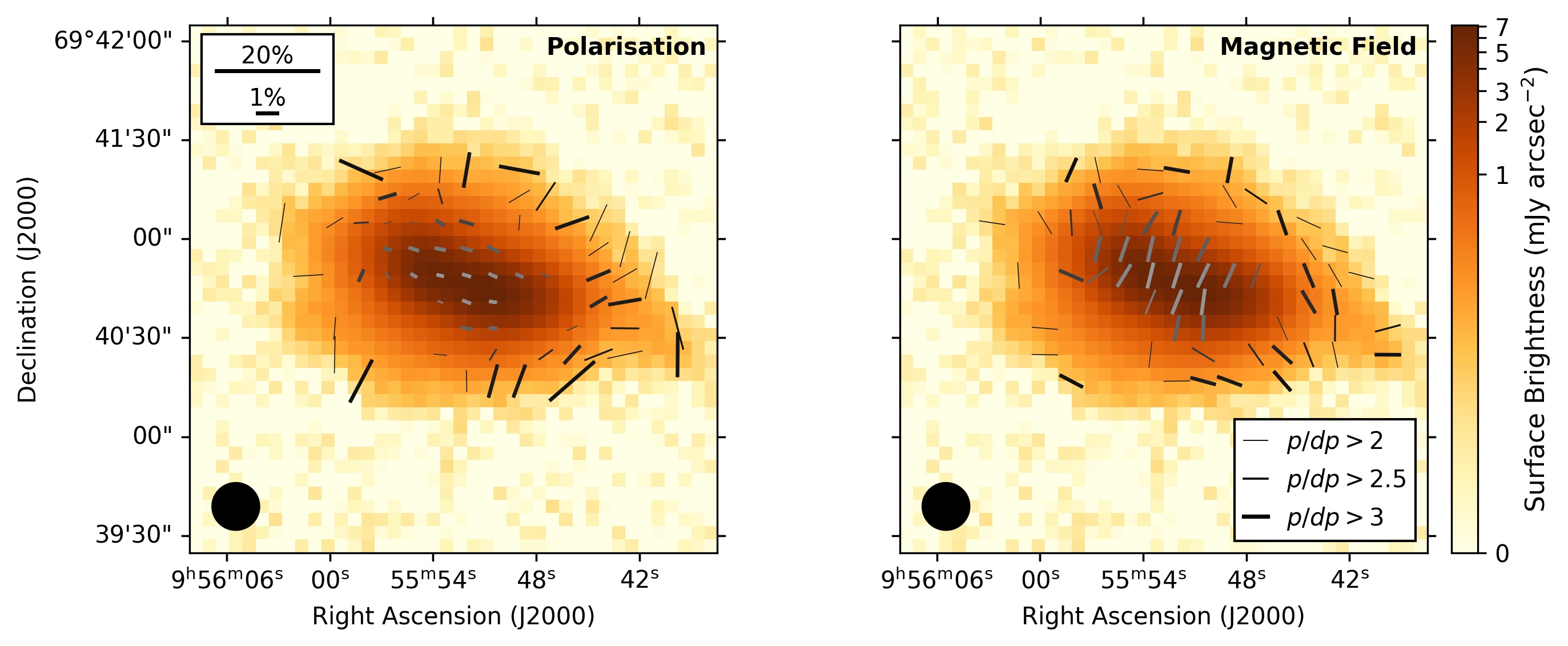}
    \caption{850$\mu$m POL-2 observations of M82.  Left panel: POL-2 polarisation vectors, with length shown in square-root scaling for clarity.  Right panel: POL-2 B-field vectors (polarisation vectors rotated by 90$^{\circ}$), shown with uniform length.  In both panels, the vectors are overlaid on 850$\mu$m Stokes $I$ emission.  Vector thickness depends on signal-to-noise in polarisation fraction.  Vector colour scaling is arbitrary and chosen for contrast against the background image.}
    \label{fig:pol2}
\end{figure*}

We observed M82 24 times between 2021 January 03 and 2021 January 29 using the POL-2 polarimeter on the SCUBA-2 camera \citep{holland2013} on the JCMT in Band 2 weather ($0.05<\tau_{225\,{\rm GHz}}<0.08$) under project code M20BP022.  Each observation consisted of a 31-minute POL-2-DAISY scan pattern.

The data were reduced using the $pol2map$\footnote{\url{http://starlink.eao.hawaii.edu/docs/sun258.htx/sun258ss73.html}} script recently added to the \textsc{Smurf} package in the $Starlink$ software suite \citep{chapin2013}.  See \citet{pattle2021} for a detailed description of the current POL-2 data reduction process. Instrumental polarisation (IP) was corrected for using the `August 2019' IP model\footnote{\url{https://www.eaobservatory.org/jcmt/2019/08/new-ip-models-for-pol2-data/}}.  The 850$\mu$m data were calibrated using a flux conversion factor (FCF) of 3159 mJy\,arcsec$^{-2}$\,pW$^{-1}$, the standard SCUBA-2 FCF of 2340 mJy\,arcsec$^{-2}$\,pW$^{-1}$ \citep{dempsey2013} multiplied by a factor of 1.35 \citep{friberg2016}.

We binned our output vector catalogue to 8-arcsec (approximately Nyquist-sampled) pixels.  The per-pixel RMS noise values in the vector catalogue were then remodelled using the $pol2noise$ script, which models map variance as the sum of three components, based on exposure time, the presence of bright sources, and residuals.  The average RMS noise in Stokes $Q$ and $U$ in the centre of the map on 8-arcsec pixels is 0.0052 mJy\,arcsec$^{-2}$ (1.2 mJy\,beam$^{-1}$).

The observed polarised intensity is given by
\begin{equation}
    PI^{\prime} = \sqrt{Q^{2} + U^{2}}.
\end{equation}
However, this quantity is biased by its defined-positive nature.  We debiased PI using the modified asympototic estimator \citep{plaszczynski2014,montier2015}:
\begin{equation}
    PI = PI^{\prime} - \frac{1}{2}\frac{\sigma^{2}}{PI^{\prime}}\left(1-e^{-\left(\frac{PI^{\prime}}{\sigma}\right)^{2}}\right),
\end{equation}
where $\sigma^{2}$ is the weighted mean of the variances $\sigma_{Q}$ and $\sigma_{U}$,
\begin{equation}
    \sigma^{2} = \frac{Q^{2}\sigma_{Q}^{2} + U^{2}\sigma_{U}^{2}}{Q^{2} + U^{2}},
\end{equation}
calculated on a pixel-by-pixel basis.  Observed polarisation fraction $p^{\prime}$ is then given by $p^{\prime} = PI^{\prime}/I$, and equivalently, debiased polarisation fraction by $p = PI/I$.  In the following analysis, we use $p$ rather than $p^{\prime}$ except where specifically stated otherwise.

Polarisation angle is given by
\begin{equation}
    \theta_{p} = 0.5\arctan(U,Q).
\end{equation}

Throughout this work we assume that dust grains are aligned with their minor axis parallel to the magnetic field direction \citep[e.g.][]{andersson2015}, and so that the plane-of-sky magnetic field direction can be inferred by rotating $\theta_{p}$ by 90$^{\circ}$.  We discuss the validity of this assumption in Section~\ref{sec:grains}, below.  We note that the polarisation angles which we detect are not true vectors, as they occupy a range in angle $0-180^{\circ}$.  We nonetheless refer to our measurements as vectors for convenience, in keeping with the general convention in the field.

\section{Results}
\label{sec:results}

The polarisation vector maps observed with POL-2 are shown in Figure~\ref{fig:pol2}.  We show all vectors with $p/dp > 2$ and $I/dI > 10$.  Vector weights in Figure~\ref{fig:pol2} show signal-to-noise ratio in $p/dp$.  It can be seen that the position angles of the $p/dp > 2$ vectors agree well with those of the $p/dp > 3$ vectors, and so we include them in our analysis.

Our results agree well with both the original SCUPOL 850$\mu$m vector map of \citet{greaves2000}, and the reprocessed map of \citet{matthews2009}.  We note that the SCUBA/SCUPOL and SCUBA-2/POL-2 systems have nothing in common apart from the JCMT dish itself.  The POL-2 and SCUPOL measurements were made using separate polarimeters, cameras, observing modes and data reduction algorithms, and so are fully independent.

\subsection{Magnetic field morphology}

We see, broadly, two behaviours: in the galactic centre, the field is perpendicular to the direction of the bar, while in the outer galaxy, the field is parallel to the spiral arm structure.

\begin{figure}
    \centering
    \includegraphics[width=0.47\textwidth]{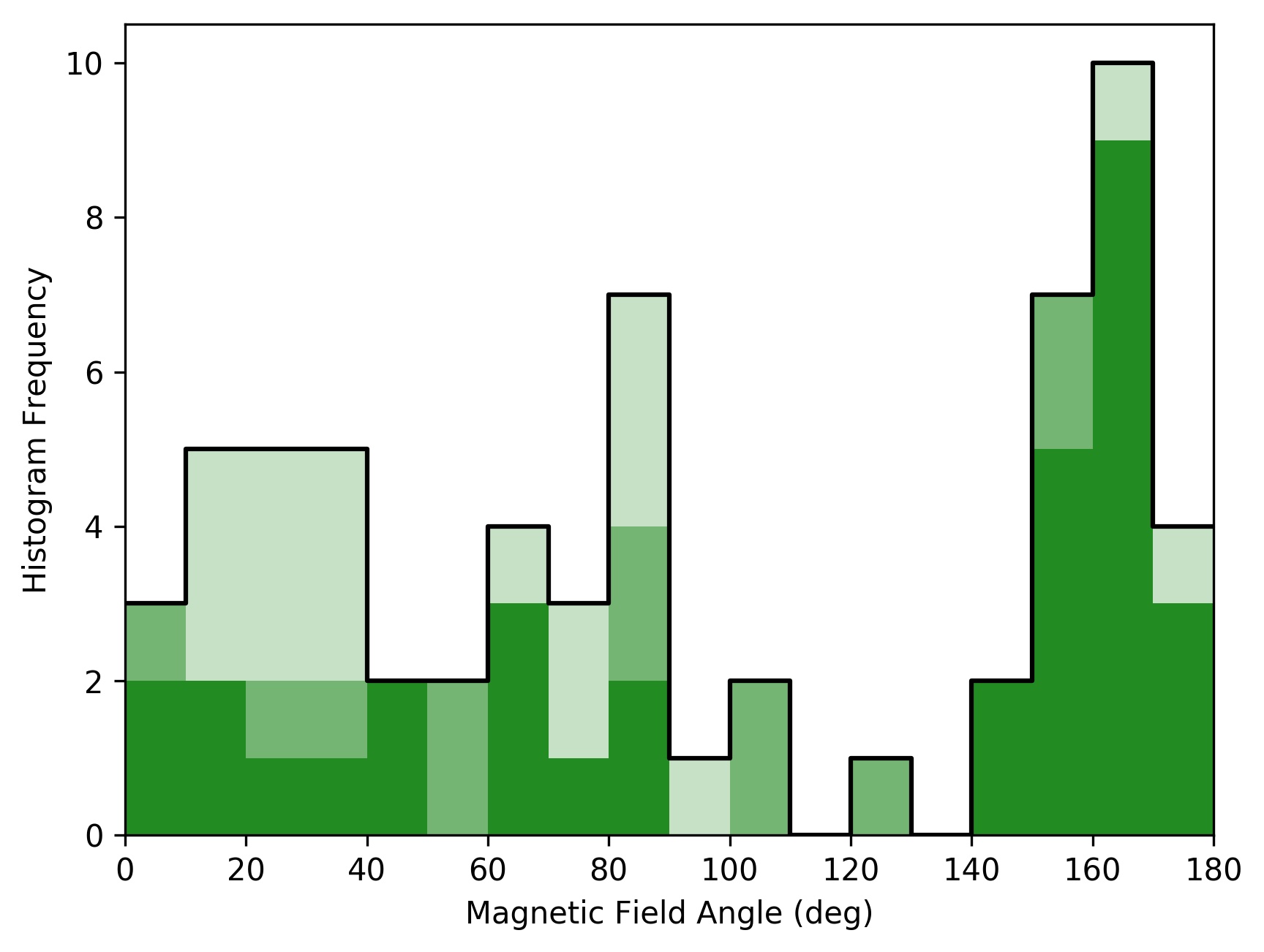}
    \includegraphics[width=0.4\textwidth]{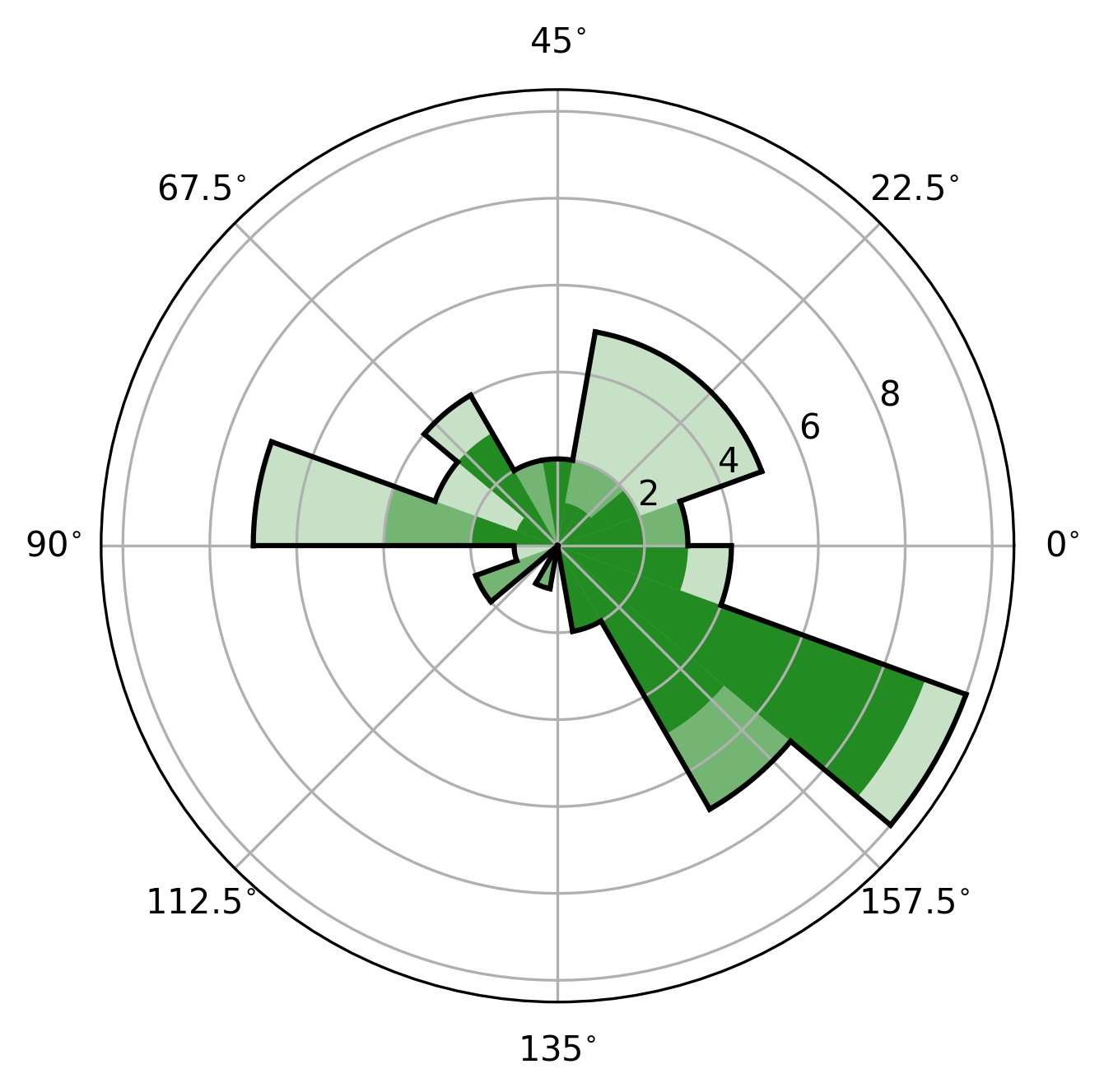}
    \caption{Histograms showing the distribution of magnetic field angles in M82.  Top panel: conventional histogram; bottom panel: circular histogram, showing the wrapping of magnetic field angle at $0^{\circ}/180^{\circ}$.  Shading denotes SNR: dark green indicates vectors with $p/dp >3$; medium green, $p/dp > 2.5$; light green, $p/dp >2$.  Note the sharp peak at $\sim 161^{\circ}$ indicating vectors tracing the poloidal field in the superwind.}
    \label{fig:angles}
\end{figure}

The distribution of magnetic field angles is shown in Figure~\ref{fig:angles}.  The field in the galactic centre can be seen as a strong peak at $161^{\circ}\pm 13^{\circ}$ E of N (circular mean value, calculated over vectors where $I > 1.1$\,mJy\,arcsec$^{-2}$), while the vectors associated with the outer galaxy occupy a broad range of angles, principally but not exclusively in the range $\sim 10-100^{\circ}$ E of N.

\begin{figure}
    \centering
    \includegraphics[width=0.47\textwidth]{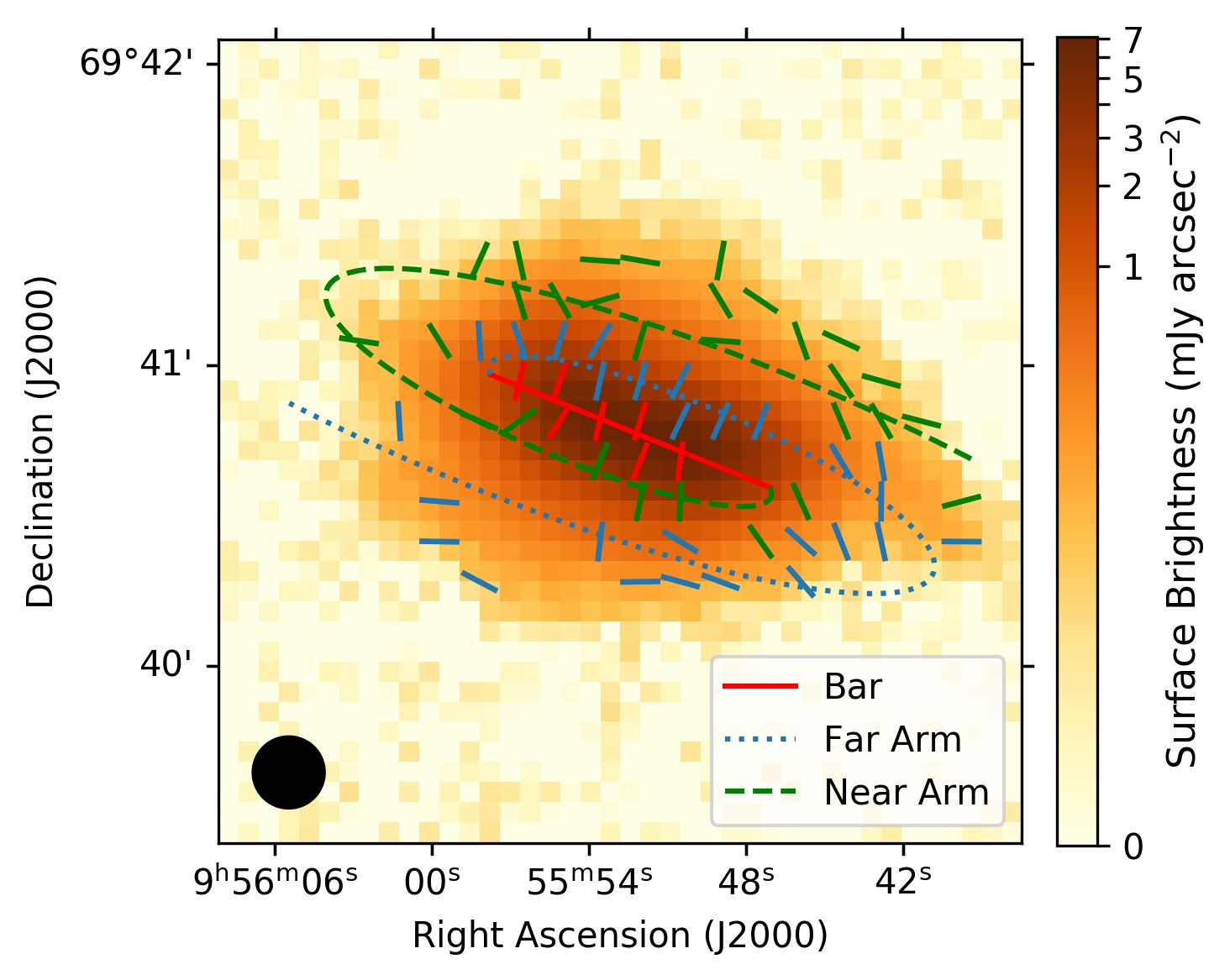}
    \includegraphics[width=0.47\textwidth]{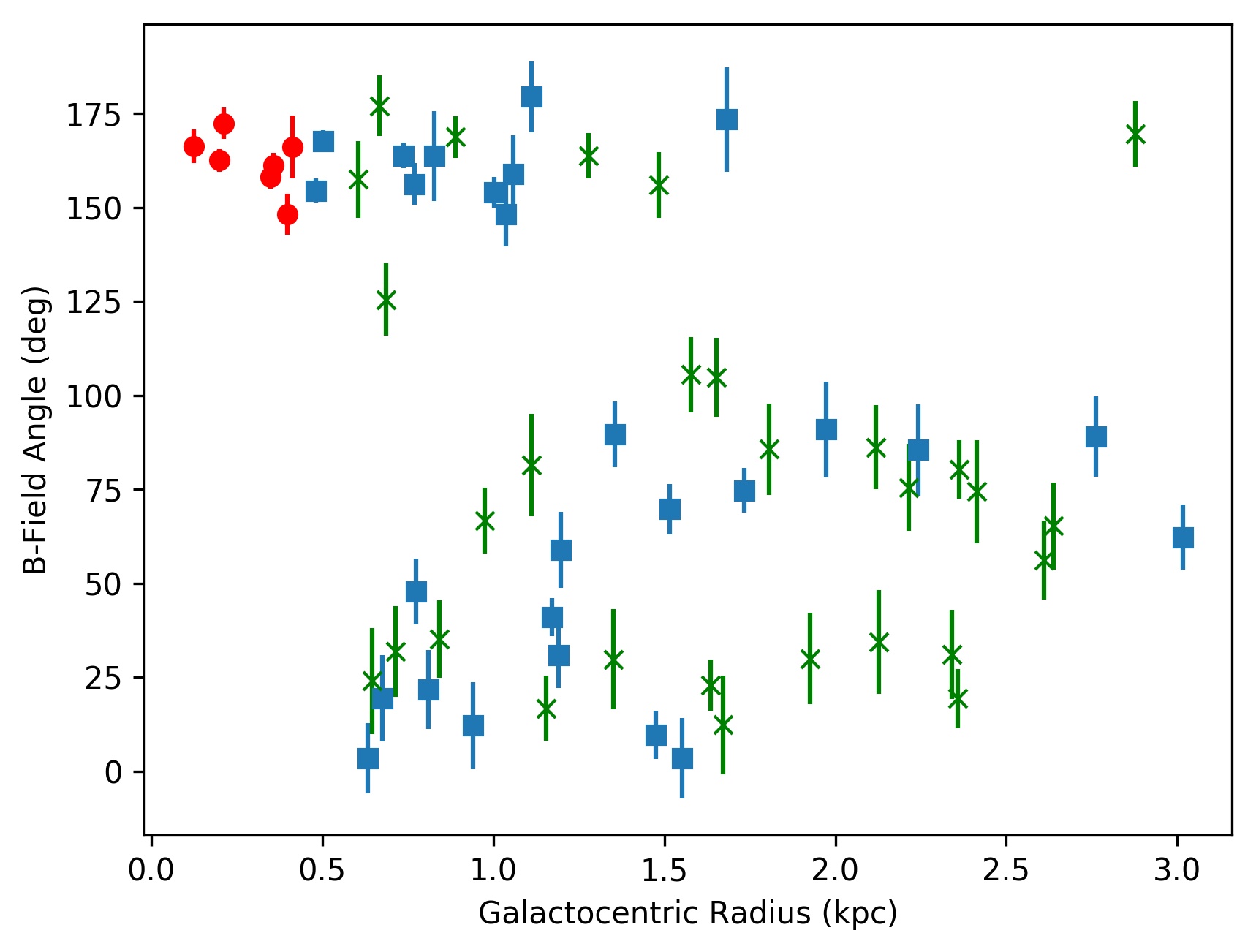}
    \includegraphics[width=0.47\textwidth]{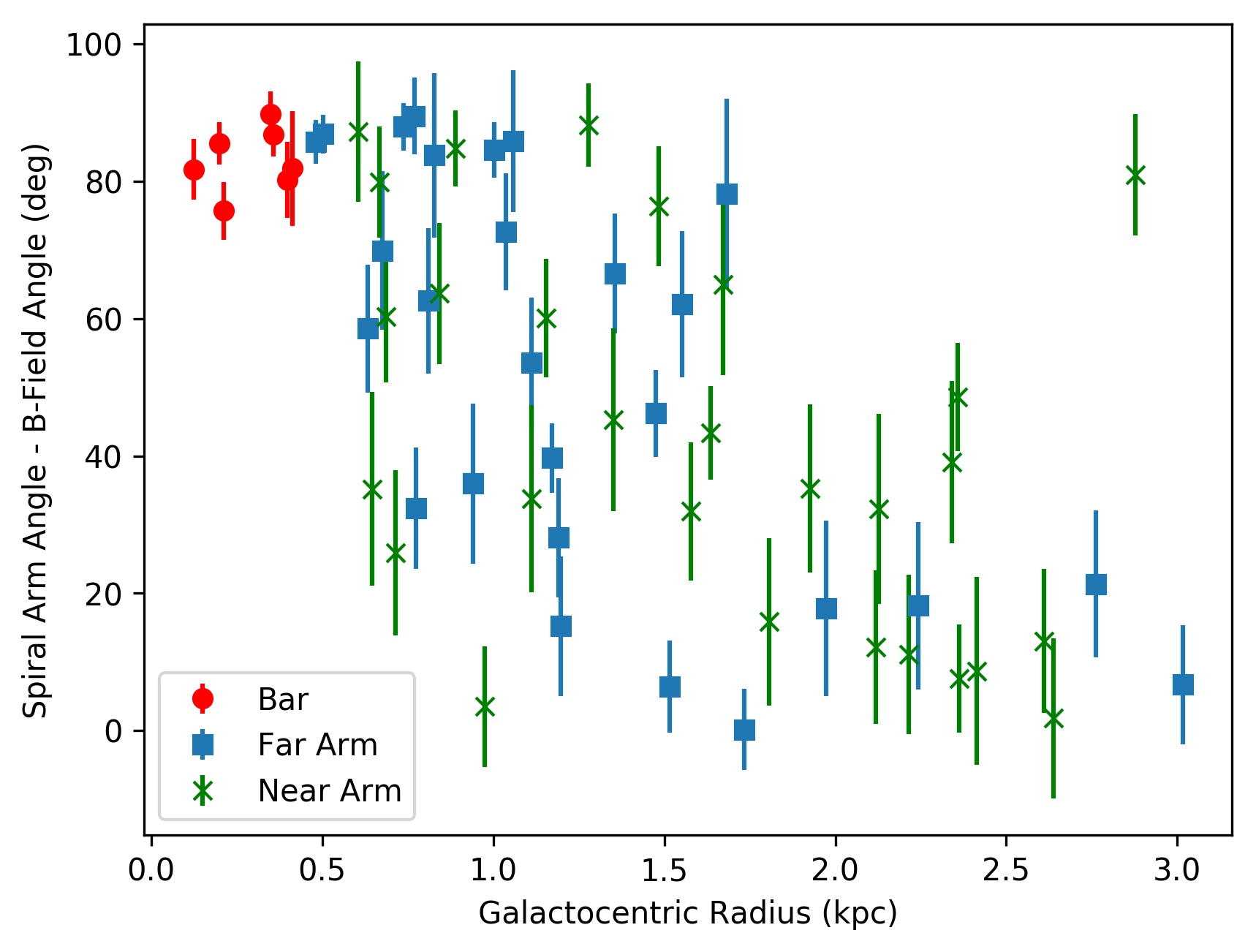}
    \caption{Comparison of magnetic field and spiral arm angles.  Top panel shows position of spiral arms \citep{mayya2005}; vectors are colour coded by their nearest spiral arm.  Middle panel: Magnetic field angle as a function of implied galactocentric radius.  Bottom panel: Difference in angle between spiral arm structure and magnetic field as a function of implied galactocentric radius.  The magnetic field appears to transition from being perpendicular to the bar in the galactic centre to being parallel to the spiral arms, or toroidal, at high galactocentric radius.}
    \label{fig:spiral}
\end{figure}

We calculated the implied galactocentric radius of each pixel if it were tracing emission from the plane of the galaxy, assuming an inclination angle of 76.9$^{\circ}$ \citep{clark2018}.  Magnetic field angle as a function of implied galactocentric radius is shown in the central panel of Figure~\ref{fig:spiral}, in which the two components of the angle distribution can be seen.

We adopted the spiral arm model of M82 proposed by \citet{mayya2005}, in which an $m=1$ logarithmic spiral arm with a pitch angle of 14$^{\circ}$ extends from each end of the 60$^{\prime\prime}$-long bar, which is offset by 4$^{\circ}$ from the 26$^{\circ}$ position angle of the galaxy \citep{telesco1991}.  These spiral arms are overlaid on our data in the top panel of Figure~\ref{fig:spiral}.  It can be seen that the poloidal $\sim 161^{\circ}$ field component extends $\sim 20^{\prime\prime}$ above and below the bar, but does not extend beyond the ends of the bar in the plane of the galaxy.  Elsewhere, the field appears to be broadly toroidal around the galactic centre, or parallel to the spiral arm structure, as discussed below.

The offset between each magnetic field vector and the angle of the spiral arm component to which it is nearest is shown in the lower panel of Figure~\ref{fig:spiral}.  We can see the transition from a perpendicular, poloidal, field pattern which dominates in the galactic centre to a field approximately parallel to the spiral arms in the outer galaxy.  With the exception of one data point at a radius of $\sim 3$\,kpc which is offset from the spiral arm direction by $\sim 80^{\circ}$ (R.A.$=09^{h}55^{m}49^{s}$, Dec.$=+69^{\circ}41^{\prime}20^{\prime\prime}.9$), the spiral-arm-aligned component emanating from the disc appears to dominate over the poloidal component beyond a galactocentric radius $\sim 2$\,kpc.  The transition occurs at an 850$\mu$m flux density $\sim 0.6$\,mJy\,arcsec$^{-2}$, at a plane-of-sky distance $\sim 20^{\prime\prime}$ above/below the bar.  Assuming an inclination of $76.9^{\circ}$ and a distance of 3.5\,Mpc, this corresponds to a height above the disc of $\sim 350$\,pc.

\subsection{Grain alignment}
\label{sec:grains}

Observations of polarised dust emission typically show a power-law dependence, $p\propto I^{-\alpha}$, where $0\leq \alpha \leq 1$ \citep{whittet2008,jones2015}.  A steeper index (higher $\alpha$) indicates either poorer grain alignment with respect to the magnetic field or more variation of the magnetic field direction along the line of sight (LOS): $\alpha = 0$ indicates that grains are consistently aligned throughout the LOS, while $\alpha = 1$ implies complete randomisation of either grain alignment or magnetic field direction along the LOS \citep{pattle2019a}.

We measured $\alpha$ using the method described by \citet{pattle2019a}, in which we assume that the underlying relationship between $p$ and $I$ can be parameterised as
\begin{equation}
    p = p_{\sigma_{QU}}\left(\frac{I}{\sigma_{QU}}\right)^{-\alpha}
    \label{eq:polfrac}
\end{equation}
where $p_{\sigma_{QU}}$ is the polarisation fraction at the RMS noise level of the data $\sigma_{QU}$, and $\alpha$ is a power-law index in the range $0 \leq \alpha \leq 1$.  We fitted the relationship between $I$ and observed non-debiased polarisation fraction $p^{\prime}$ with the mean of the Ricean distribution of observed values of $p$ which would arise from equation~\ref{eq:polfrac} in the presence of Gaussian RMS noise $\sigma_{QU}$ in Stokes $Q$ and $U$:
\begin{equation}
    p^{\prime}(I) = \sqrt{\frac{\pi}{2}}\left(\frac{I}{\sigma_{QU}}\right)^{-1}\mathcal{L}_{\frac{1}{2}}\left(-\frac{p_{\sigma_{QU}}^{2}}{2}\left(\frac{I}{\sigma_{QU}}\right)^{2(1-\alpha)}\right).
    \label{eq:rmfit}
\end{equation}
where $\mathcal{L}_{\frac{1}{2}}$ is a Laguerre polynomial of order $\frac{1}{2}$.  See \citet{pattle2019a} for a derivation of this result.  We restricted our data set to the central 3-arcminute diameter region over which exposure time, and so RMS noise, is approximately constant \citep{friberg2016}.

\begin{figure}
    \centering
    \includegraphics[width=0.47\textwidth]{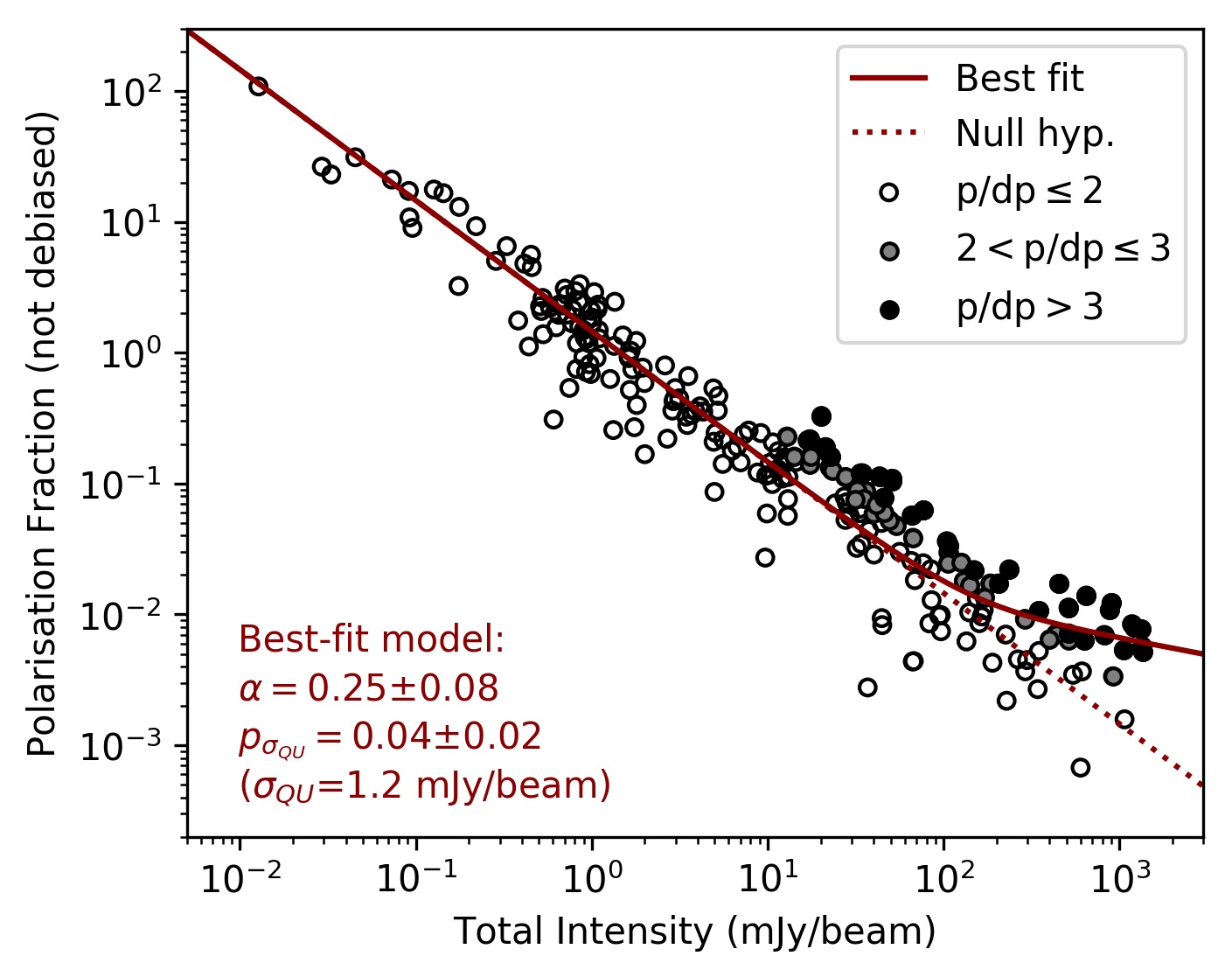}
    \caption{Non-debiased polarisation fraction as a function of Stokes $I$ intensity, fitted with a single-power-law distribution and a Ricean noise model, as described in the text.  All data points in the central 3-arcmin-diameter region of the image are shown and fitted; those at $p/dp > 2$ are shaded in grey; those at $p/dp > 3$ are shaded in black.  The best-fit model, with a power-law index $\alpha=0.25\pm0.08$, determined by fitting Equation~\ref{eq:rmfit} to the data, is shown as a solid red line.  The behaviour in the absence of true polarised signal, $p^{\prime} = \sqrt{\pi/2}(I/\sigma_{QU})^{-1}$, is shown as a dashed red line.}
    \label{fig:p_vs_I}
\end{figure}

The relationship between $p^{\prime}$ and $I$ in M82 is shown in Figure~\ref{fig:p_vs_I}.  By fitting Equation~\ref{eq:rmfit} to the data, we measure a best-fit index of $\alpha = 0.25\pm 0.08$.  This suggests that in our observations of M82, grain alignment remains quite good and that a single average field direction along the LOS dominates the 850$\mu$m emission at most locations, implying a sharp transition from the poloidal component to the orthogonal toroidal component dominating the emission profile above and below the bar.  We see hints of a line of null polarisation at $\sim 25^{\prime\prime}$ above and below the bar, perhaps delineating the locations where the wind and disc dust components contribute similar amounts of polarised 850$\mu$m emission.

\section{Discussion}
\label{sec:discussion}

We posit that the polarisation geometry observed at 850$\mu$m can easily be reconciled with observations at other wavelengths if it traces the poloidal magnetic field in the central starburst at small galactocentric radii, and a toroidal/spiral-arm-aligned field in the disc of M82 at large galactocentric radii.  In Figure~\ref{fig:spiral} we show that the magnetic field direction transitions from being broadly perpendicular to the bar in the galactic centre to being broadly parallel to the spiral arms at larger galactocentric radii.

Many previous studies have suggested the existence of a two-component magnetic field in M82 \citep[c.f.][Figure 5]{jones2000}.  Observations of other spiral galaxies (e.g. M51; \citealt{jones2020}) show magnetic fields running parallel to the spiral arms, in keeping with the $\alpha\omega$-dynamo model.  We note that at the signal-to-noise ratio that we achieve in the outer parts of M82, it is difficult to definitively distinguish between a field generically toroidal around the galactic centre and one running along the spiral arms.  However, both theory and previous observation leads us to expect fields to be parallel to spiral arms where such arms exist \citep{beck1996}.

\begin{figure}
    \centering
    \includegraphics[width=0.47\textwidth]{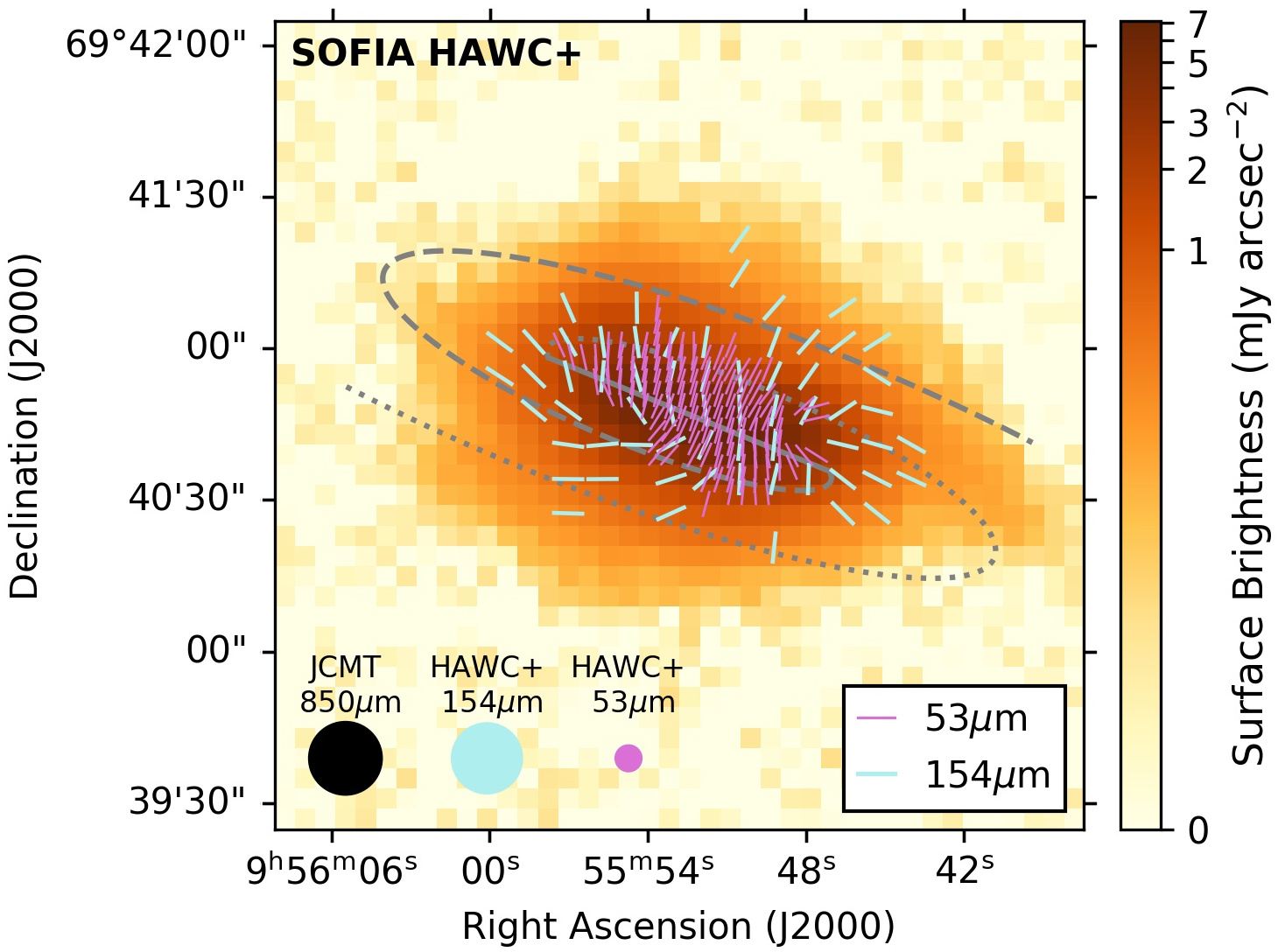}
    \caption{HAWC+ 53$\mu$m (thin magenta) and 154$\mu$m (thick light blue) magnetic field vectors \citep{jones2019}, overlaid on 850$\mu$m Stokes $I$ emission.  \citet{mayya2005} spiral arm model is shown in light grey.}
    \label{fig:hawc}
\end{figure}

HAWC+ 53$\mu$m and 154$\mu$m vectors \citep{jones2019}, are shown in Figure~\ref{fig:hawc}.  We see good agreement between the POL-2 and HAWC+ results in the centre of the galaxy, with all three wavelengths appearing to trace the poloidal field component.  While the HAWC+ 53$\mu$m measurements do not extend far beyond the central bar, the 154$\mu$m measurements extend considerably further.  The POL-2 850$\mu$m and HAWC+ 154$\mu$m results agree at the ends of the bar, where we do not expect the outflow to be launching material.  However, the two measurements disagree in the outer galaxy both above and below (particularly above) the galactic plane.

The fact that HAWC+ observations trace the galactic wind indicates that large amounts of dust are entrained in the galactic outflow \citep{jones2019}.  Based on their 53$\mu$m observations, \citet{jones2019} infer that the poloidal field geometry extends up to 350\,pc above and below the plane, while the outflow itself extends at least 11\,kpc from the plane \citep{devine1999}.  

We see the poloidal field extending to $\sim 350$\,pc above and below the plane, in very good agreement with the extent over which the HAWC+ 53$\mu$m emission is observed.  However, the 154$\mu$m-inferred field geometry continues to be poloidal to significantly larger distances above the plane, and the superwind field lines have been extrapolated to extend into the intergalactic medium \citep{lopezrodriguez2021}.  This suggests that at 850$\mu$m, we are observing the poloidal magnetic field in the central starburst region.  Beyond this region, our observations trace the magnetic field in the cold, high-column-density dust of the galactic plane, while 154$\mu$m observations trace hot, low-column-density dust entrained in the superwind, in which the dust mass scale height is $1.4\pm 0.3$\,kpc \citep{leroy2015}.
The change in behaviour at $\sim$350\,pc above the plane is consistent with models of M82 that call for the breakout from the starburst region to the beginning of the superwind to take place at around this vertical scale \citep[e.g.][]{heckman1990,martini2018}.

\section{Conclusions}
\label{sec:summary}

We have observed the starburst galaxy M82 in 850$\mu$m polarised light using the POL-2 polarimeter on the James Clerk Maxwell Telescope (JCMT). 
Our observations trace a poloidal magnetic field in the M82 central starburst region to heights $\sim 350$\,pc above the plane of the galaxy, but trace a field in the disc parallel to the spiral arms at galactocentric radii $\gtrsim 2$\,kpc, in good agreement with predictions for a starbursting spiral galaxy.
We see a significant discrepancy between POL-2 850$\mu$m and HAWC+ 154$\mu$m measurements in the outer galaxy, where the HAWC+ measurements trace hot, low-column-density dust entrained by the superwind.  Observations across the submillimetre/far-infrared regime are thus necessary to disentangle the multiple magnetic field components of starburst galaxies.

\section*{Acknowledgements}

We thank Terry J. Jones for providing us with the HAWC+ vector catalogues.  The James Clerk Maxwell Telescope is operated by the East Asian Observatory on behalf of The National Astronomical Observatory of Japan; Academia Sinica Institute of Astronomy and Astrophysics; the Korea Astronomy and Space Science Institute; Center for Astronomical Mega-Science (as well as the National Key R\&D Program of China with No. 2017YFA0402700). Additional funding support is provided by the Science and Technology Facilities Council of the United Kingdom and participating universities and organisations in the United Kingdom, Canada and Ireland.  Additional funds for the construction of SCUBA-2 were provided by the Canada Foundation for Innovation.  This research made use of Astropy \citep{astropy:2013, astropy:2018}.   The authors wish to recognise and acknowledge the very significant cultural role and reverence that the summit of Maunakea has always had within the indigenous Hawaiian community.  We are most fortunate to have the opportunity to conduct observations from this mountain.

\section*{Data Availability}

The reduced data used in this analysis are available at [CADC DOI to be inserted].  The raw data are available in the Canadian Astronomy Data Centre archive under project code M20BP022.


\bibliographystyle{mnras}



\bsp	
\label{lastpage}
\end{document}